\documentclass[twocolumn,pra,showpacs,floatfix]{revtex4}
\usepackage{graphics}

\begin{document}

\title{Optical phase-space reconstruction of mirror motion at the attometer level}

\author{T. Briant}
\author{P.F. Cohadon}
\author{M. Pinard}
\author{A. Heidmann}

\affiliation{Laboratoire Kastler Brossel, Case 74, 4 place Jussieu, F75252
Paris Cedex 05, France}
\thanks{Unit\'{e} mixte de recherche du Centre National de la Recherche
Scientifique, de l'Ecole Normale Sup\'{e}rieure et de l'Universit\'{e} Pierre
et Marie Curie} \homepage{www.spectro.jussieu.fr/Mesure}

\date{20 September 2002}

\begin{abstract}
We describe an experiment in which the quadratures of the position
of an harmonically-bound mirror are observed at the attometer
level. We have studied the Brownian motion of the mirror, both in
the free regime and in the cold-damped regime when an external
viscous force is applied by radiation pressure. We have also
studied the thermal-noise squeezing when the external force is
parametrically modulated. We have observed both the $50\;\%$
theoretical limit of squeezing at low gain and the parametric
oscillation of the mirror for a large gain.
\end{abstract}

\pacs{42.50.Lc, 05.40.Jc, 04.80.Nn}

\maketitle

\section{Introduction}

\label{Sec_Intro}

Optomechanical coupling, that is the cross-coupling between the
motion of a mirror and a laser field reflected upon it,\ first
appeared in the context of interferometric gravitational-wave
detection \cite {Bradaschia90,Abramovici92} with the existence of
the so-called Standard Quantum Limit
\cite{Caves81,Jaekel90,Braginsky92}. The unique sensitivity of
interferometric techniques has since then been used for other
high-sensitivity measurements, such as AFM \cite{Rugar89} or
optical transducers \cite{Stephens93,Conti98}. The field has
recently encountered much attention and has gained a life of its
own in the quantum optics community: several schemes involving a
cavity with a movable mirror have been proposed either to create
non-classical states of both the radiation field \cite
{Fabre94,Mancini94} and of the motion of the mirror \cite
{Bose97,Bose99,Mancini02}, or to perform QND measurements
\cite{Heidmann97}. Recent progress in low-noise laser sources and
low-loss mirrors have made the field experimentally accessible
\cite{Hadjar99,Cohadon99,Tittonen99}.

A recent trend in quantum optics is to fully reconstruct the
quantum state of either a mode of the radiation field through the
quantum tomography technique \cite{Vogel89,Smithey93} or a trapped
atom \cite{Leibfried96}, but to our knowledge, no such experiment
has yet been performed at the quantum level on a mechanical
oscillator.

In this paper, we present an experiment which completes the
analysis of the motion of a plano-convex mirror given in
\cite{Hadjar99,Cohadon99,Pinard01}, reconstructing the phase-space
distribution of the motion through the simultaneous classical
measurement of both quadratures of the mirror position. The
technique is applied to a variety of states of motion: Brownian
motion and its cold-damped and squeezed counterparts.

In Sec. \ref{Sec_PhaseSpace} we present the experimental setup used to
monitor the motion of the mirror and we describe how the motion in
phase-space can be reconstructed. In Sec. \ref{Sec_Brownian} we consider the
case of a free mirror at thermal equilibrium and we compare our experimental
results to predictions of the fluctuations-dissipation theorem \cite{Callen51}%
. The experimental setup has also been modified to include an external
viscous force applied to the mirror, and a similar analysis is given in Sec.
\ref{Sec_ColdDamping} in the corresponding case of the cold-damped regime.
Our results demonstrate that a new equilibrium is obtained with a reduction
of thermal noise.

In Sec. \ref{Sec_Param} we present an experiment of parametric
amplification of the thermal noise \cite{Rugar91}, below the
oscillation threshold. The experimental setup is modified in order
to modulate the strength of the viscous force, and squeezing of
the thermal noise has been observed. The observation of parametric
oscillations of the mirror is finally described in Sec.
\ref{Sec_ParamOsc}.

\section{Evolution in phase-space}

\label{Sec_PhaseSpace}

The mirror motion is monitored by an optomechanical displacement
sensor. It relies on the sensitivity of the phase of a reflected
light beam to mirror displacements. Monitoring such a phase-shift
allows to reconstruct the dynamical evolution of the mirror.\ This
phase-shift can be induced by various kinds of mirror motions,
either external \cite{Tittonen99,Dorsel83} or internal
\cite{Hadjar99,Bondu95,Gillespie95}.\ The former is important for
suspended mirrors since the excitation of pendulum modes of the
suspension system leads to global displacements of the mirror.\
The latter corresponds to deformations of the mirror surface due
to the excitation of internal acoustic modes of the substrate.
These various degrees of freedom have however different resonance
frequencies and one can select the mechanical response of a
particular mode by using a bandpass filter.

In our experiment we detect only frequencies around the fundamental acoustic
resonance of the mirror so that the mechanical response is mainly ruled by
the behaviour of this particular internal mode.\ The mirror motion can then
be approximated as the one of a single harmonic oscillator characterized by
its resonance frequency $\Omega _{M}$, its quality factor $Q$, and its mass $%
M$. This mass actually corresponds to an effective mass which
describes the effective motion of the mirror as seen by the light,
that is the deformation of the mirror surface averaged over the
beam spot \cite {Bondu95,Gillespie95,Pinard99}.

The temporal evolution of the mirror position $x\left( t\right) $
is given by linear response theory \cite{Landau58}.\ Assuming that
a force $F$ is applied to the mirror, the displacement $x\left[
\Omega \right] $ in Fourier space at frequency $\Omega $ is
related to the force by
\begin{equation}
x\left[ \Omega \right] =\chi \left[ \Omega \right] F\left[ \Omega \right] ,
\label{Equ_x}
\end{equation}
where $\chi \left[ \Omega \right] $ is the mechanical
susceptibility of the mirror.\ For a viscously-damped harmonic
oscillator it has a Lorentzian behaviour given by
\begin{equation}
\chi \left[ \Omega \right] =\frac{1}{M\left( \Omega _{M}^{2}-\Omega
^{2}-i\Gamma \Omega \right) },  \label{Equ_Chi}
\end{equation}
where $\Gamma =\Omega _{M}/Q$ is the damping rate
\cite{CommentDamping}.

Depending on the nature of the applied force $F$, the spectrum of the mirror
motion will typically be peaked around the mechanical resonance frequency $%
\Omega _{M}$ over a frequency span $\Gamma $, which is much smaller than $%
\Omega _{M}$ for a high-$Q$ harmonic oscillator.\ To study the temporal
evolution of the motion in phase-space, it is preferable to describe the
motion in the rotating frame in order to remove the intrinsic oscillatory
dependence with time. This leads to the two quadratures $X_{1}$ and $X_{2}$,
defined by
\begin{equation}
x\left( t\right) =X_{1}\left( t\right) \cos \left( \Omega _{M}t\right)
+X_{2}\left( t\right) \sin \left( \Omega _{M}t\right) .  \label{Equ_quad}
\end{equation}
This equation yields the expression of the two quadratures in Fourier space,
\begin{eqnarray}
X_{1}\left[ \omega \right] &=&x\left[ \Omega _{M}+\omega \right] +x\left[
-\Omega _{M}+\omega \right] ,  \label{Equ_X1} \\
X_{2}\left[ \omega \right] &=&-i\left( x\left[ \Omega _{M}+\omega \right] -x%
\left[ -\Omega _{M}+\omega \right] \right) .  \label{Equ_X2}
\end{eqnarray}
Quadratures $X_{1}\left( t\right) $ and $X_{2}\left( t\right) $ vary very
slowly with time (over a $1/\Gamma $ timescale), and $\omega $ which
represents the frequency mismatch $\Omega -\Omega _{M}$ between the analysis
frequency and the mechanical resonance frequency is always considered small
as compared to $\Omega _{M}$.

Expressions of the quadratures in presence of an applied force can be
derived from eq.\ (\ref{Equ_x}) by inserting the expressions of $x\left[
\Omega _{M}+\omega \right] $ and $x\left[ -\Omega _{M}+\omega \right] $ in
the definitions of $X_{1}\left[ \omega \right] $ and $X_{2}\left[ \omega %
\right] $. We find after simplification,
\begin{eqnarray}
X_{1}\left[ \omega \right] &=&-\frac{1}{2M\Omega _{M}}\left( \frac{1}{%
-i\omega +\Gamma /2}\right) F_{2}\left[ \omega \right] ,  \label{Equ_X1/F} \\
X_{2}\left[ \omega \right] &=&\frac{1}{2M\Omega _{M}}\left( \frac{1}{%
-i\omega +\Gamma /2}\right) F_{1}\left[ \omega \right] ,  \label{Equ_X2/F}
\end{eqnarray}
where we have introduced the two quadratures $F_{1}$ and $F_{2}$
of the applied force, defined in Fourier space by similar
expressions as for $X_{1}$ and $X_{2}$ (eqs.\ \ref{Equ_X1} and
\ref{Equ_X2}). Both position quadratures have a Lorentzian
response centered at zero frequency and of width $\Gamma $.

\begin{figure}[h]
\resizebox{6 cm}{!}{\includegraphics{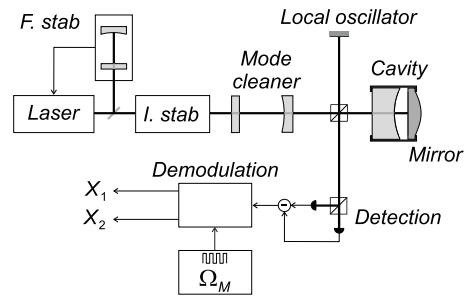}}
\caption{Experimental setup. A plano-convex mirror is used as the end mirror of a high-finesse
cavity. A frequency and intensity-stabilized laser beam is sent into the
cavity and the phase of the reflected beam is measured by a homodyne
detection. A demodulation system working at the mechanical resonance
frequency $\Omega _{M}$ extracts both quadratures of the signal.}
\label{Fig_Setup}
\end{figure}

The harmonic oscillator we consider throughout the paper is the
high-$Q$ fundamental acoustic mode of a mirror coated on a
plano-convex resonator made of fused silica. The resonator is
$1.5\ mm$ thick at the center with a diameter of $14\ mm$ and a
curvature radius of the convex side of $100\ mm$. Such dimensions
lead to a resonance frequency of the fundamental mode of the order
of $2\ MHz$. The oscillator's parameters have the following values
\cite {Hadjar99},
\begin{equation}
\Omega _{M}=2\pi \times 1859\ kHz,\;M=230\ mg,\;Q=44000.  \label{Equ_Params}
\end{equation}
The mirror coated on the flat side of the resonator is used as the end
mirror of a single-ended Fabry-Perot cavity with a {\it Newport high-finesse
SuperMirror} as input mirror (figure \ref{Fig_Setup}).\ The whole provides a
$1\ mm$-long cavity with an optical finesse ${\cal F}=37000$. The light beam
entering the cavity is provided by a frequency-stabilized titane-sapphire
laser working at $\lambda =810\ nm$. The light beam is also
intensity-stabilized and spatially filtered by a mode cleaner.

Near an optical resonance of the cavity the intracavity intensity shows an
Airy peak when the cavity length is scanned through the resonance, and the
phase of the reflected field is shifted by $2\pi $. The slope of this
phase-shift strongly depends on the cavity finesse and for a displacement $%
\delta x$ of the end mirror one gets at resonance a phase-shift $\delta
\varphi $ on the order of
\begin{equation}
\delta \varphi \simeq 8{\cal F}\frac{\delta x}{\lambda }+\delta \varphi _{n},
\label{Equ_Phi}
\end{equation}
where $\lambda $ is the optical wavelength and $\delta \varphi _{n}$ the
phase noise of the reflected beam.\ At high frequency, all technical noises
can be suppressed and the phase noise corresponds to the quantum noise of
the incident beam which is inversely proportional to the square root of the
light power.

The phase of the reflected field is monitored by a homodyne detection
working at the quantum level. For a $100\;\mu W$ incident laser beam, we
have shown that the sensitivity to mirror displacements is limited by the
quantum noise of light and it has been measured to
\begin{equation}
\delta x_{\min }=2.8\;10^{-19}\ m/\sqrt{Hz},  \label{Equ_dxmin}
\end{equation}
at an analysis frequency of $2\ MHz$ \cite{Hadjar99}.

\begin{figure}[h]
\resizebox{6.5 cm}{!}{\includegraphics{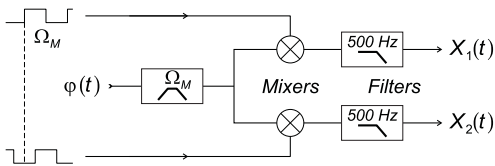}}
\caption{Demodulation of the
position signal: the signal $\varphi (t)$ given by the homodyne
detection is filtered and mixed with two $\pi /2$-dephased square
signals at the mechanical resonance frequency $\Omega _{M}$.
Low-pass filters are used to extract the low-frequency component
of the quadratures.} \label{Fig_Demodul}
\end{figure}

To extract both quadratures of the mirror motion we use a demodulation
system at frequency $\Omega _{M}$ (figure \ref{Fig_Demodul}). A bandpass
filter of width $10$ $kHz$ first selects frequencies around the fundamental
resonance frequency in the signal given by the homodyne detection.\ The
signal is then mixed with two square signals at frequency $\Omega _{M}$, one
dephased by $\pi /2$ with respect to the other one. Harmonics of the output
signals are cancelled out by two electronic low-pass filters. The filters
also have to cancel the signal due to other acoustic modes of both the
resonator and the coupling mirror. The expected width of the fundamental
mode being of the order of $43\ Hz$ and the closest mode $3\ kHz$ above, we
use second-order filters with a $500\ Hz$ cut-off frequency. We have checked
the transfer function of each filter in order to ensure that the dissymmetry
between both channels is less than $1\%$.

Note that $X_{1}$ and $X_{2}$ are conjugate quantum observables
and therefore cannot be measured simultaneously at the quantum
level. This is not in contradiction with our experimental setup
since we extract the position quadratures $X_{1}$ and $X_{2}$ from
the phase $\varphi $ of the reflected field which also contains
the quantum phase noise $\delta \varphi _{n}$ of light (eq.\
\ref{Equ_Phi}). A measurement of the quadratures is thus
contaminated by the quantum noise of light so that it is not
possible to measure simultaneously both quadratures at the quantum
level.\ For this paper this does not constitute a limitation since
the signals studied in next sections (thermal noise and external
force) are way above the Standard Quantum Limit imposed by the
measurement and back-action noises \cite
{Caves81,Jaekel90,Braginsky92}.

\section{Observation of Brownian motion}

\label{Sec_Brownian}

We present in this section the results obtained when the mirror is
free (no external force) and at room temperature.\ The thermal
equilibrium can be described as the result of a coupling with a
thermal bath via a Langevin force $F_{T}$ applied on the mirror.\
The resulting Brownian motion is the response of the acoustic mode
to this force according to eq. (\ref{Equ_x}). The
fluctuations-dissipation theorem relates the spectrum $S_{T}$ of
the Langevin force to the dissipative part of the mechanical
susceptibility \cite {Callen51},
\begin{equation}
S_{T}\left[ \Omega \right] =-\frac{2k_{B}T}{\Omega }Im\left( 1/\chi \left[
\Omega \right] \right) ,  \label{Equ_ST}
\end{equation}
where $k_{B}$ is the Boltzmann constant and $T$ the temperature of the
thermal bath.

The sensitivity of our experiment is high enough to detect the
Brownian motion.\ We have already observed the noise spectrum of
the mirror displacement, which has a Lorentzian shape centered at
the mechanical resonance frequency $\Omega _{M}$ and a width
$\Gamma $ \cite{Hadjar99}.\ The height of the peak is at least 4
orders of magnitude larger than the sensitivity defined by the
quantum phase noise of the reflected beam (eq. \ref{Equ_dxmin}).

\begin{figure}[h]
\resizebox{5 cm}{!}{\includegraphics{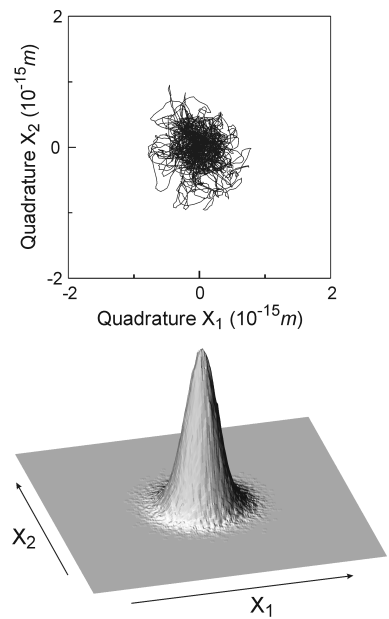}} \caption{Brownian motion
of the plano-convex mirror in the $\left(X_1,X_2\right)$ phase-space. Full
scale corresponds to $\pm 2$ $10^{-15}$ $m$. Top: temporal acquisition over
$500$ $ms$. Bottom: histogram of the distribution acquired over 10 minutes.
The number of cells is $256\times 256$, the limit of sensitivity (quantum
phase noise of light) corresponding approximately to 1 cell.}
\label{Fig_thermique}
\end{figure}

We have observed the temporal evolution of the Brownian motion by sending
the outputs of the demodulation system in a digital oscilloscope.\ The
resulting traces of the two quadratures $X_{1}$ and $X_{2}$ are plotted in
the upper curve of figure \ref{Fig_thermique} for an acquisition time of $%
500\ ms$. The temporal trajectory in phase-space appears as a random walk
around the center $\left( X_{1}=0,X_{2}=0\right) $.

This plot can be compared to the Brownian motion of a mirror coated on a
torsion oscillator, as observed in ref. \cite{Tittonen99} by a similar
optical interferometry technique. The larger density of points and the more
circular shape of our results is due to the larger ratio between the
acquisition time $\tau _{acq}$ and the evolution time ($\simeq 2\pi /\Gamma $%
) of the Brownian motion in phase-space. In our experiment, $\tau
_{acq}=500\ ms$ and $\Gamma /2\pi =43\ Hz$, so that the phase space is
mapped approximately $20$ times. In the case of the torsion oscillator, as
the mechanical resonance frequency is two orders of magnitude lower, $\Gamma
/2\pi \simeq 20\ mHz$ and the phase space is mapped just once even for a $%
\tau _{acq}=80\ s$ acquisition time. This explains why results in
ref. \cite {Tittonen99} look similar to the Brownian motion of a
free particle: ergodicity is irrelevant on such short timescales.

Our results are more likely to be compared to the thermomechanical
noise observed with an AFM cantilever \cite{Rugar91}, although the
simpler two-wave
technique used in this work yielded a lower sensitivity, on the order of $%
10^{-12}\ m$. In this experiment the resonance frequency also lay in the $%
20\ kHz$ range but the lower mechanical quality factor of the cantilever
drastically lowered the timescale of the Brownian motion, enabling to
display results comparable to ours.

We have calibrated the observed displacements by using a frequency
modulation of the laser beam.\ As shown by eq.\ (\ref{Equ_Phi}) we measure a
displacement with a reference corresponding to the optical wavelength.\ A
displacement $\Delta x$ of the mirror is thus equivalent to a frequency
variation $\Delta \nu $ of the laser related to $\Delta x$ by
\begin{equation}
\frac{\Delta \nu }{\nu }=\frac{\Delta x}{L},  \label{Equ_dnu}
\end{equation}
where $\nu $ is the optical frequency and $L$ the cavity length.\ Plots of
figure \ref{Fig_thermique} correspond to a full scale of the oscilloscope of $%
\pm 100$ $mV$. We have applied a frequency modulation of calibrated
amplitude $\Delta \nu =200$ $Hz$ and obtained a deviation in phase-space of $%
27$ $mV$. From eq.\ (\ref{Equ_dnu}) this corresponds to an equivalent
displacement $\Delta x=5.4$ $10^{-16}$ $m$, so that $1$ $mV$ is equivalent
to $2$ $10^{-17}$ $m$ and full scale in figure \ref{Fig_thermique}
corresponds to $\pm 2$ $10^{-15}$ $m$.

This calibration also allows to determine the sensitivity of the measurement
in phase-space. The quantum phase noise of light induces a rms voltage noise
at the output of the demodulation system of $0.86$ $mV$.\ The smallest
observable displacement in phase-space is then
\begin{equation}
\Delta x_{\min }=1.7\;10^{-17}\ m,  \label{Equ_DXmin}
\end{equation}
and corresponds to a few attometers. One can relate this sensitivity to the
one expressed in term of spectral amplitude (eq.\ \ref{Equ_dxmin}).\ Since
the phase noise of the reflected beam is a white noise on the frequency
scale of the measurement, the minimum measurable displacement $\Delta
x_{\min }$ in phase-space is equal to the spectral sensitivity $\delta
x_{\min }$ integrated over the measurement bandwidth, that is
\begin{equation}
\Delta x_{\min }^{2}=\left( \delta x_{\min }\right) ^{2}\int \left| H\left[
\omega \right] \right| ^{2}d\omega ,  \label{Equ_Dx2}
\end{equation}
where $H\left[ \omega \right] $ is the transfer function of the low-pass
filters at the output of the demodulation system. From the value of the
spectral sensitivity (eq. \ref{Equ_dxmin}) and the characteristics of the
filter (second-order filter with cut-off frequency $460$ $Hz$ and quality
factor $2200$), one gets
\begin{equation}
\Delta x_{\min }\simeq 1.65\;10^{-17}\ m,
\end{equation}
in excellent agreement with the measured value (eq.\ \ref{Equ_DXmin}). As
shown by eq. (\ref{Equ_Dx2}) the sensitivity depends on the frequency
cut-off of the filters and can be increased with a smaller measurement
bandwidth.

Acquiring data during a longer time allows to reconstruct the distribution
in phase-space. The full voltage scale at the output of the demodulation
system is divided into $256$ cells so that one cell approximately
corresponds to the limit of sensitivity $\Delta x_{\min }$. We then perform
an histogram by accumulating the temporal traces of the quadratures in the $%
256\times 256$ cells.\ For an acquisition time of $10$ minutes the total
number of points delivered by the digital oscilloscope is of the order of $6$
$10^{5}$ and the phase-space is mapped approximately $25000$ times. One then
gets a good statistics for the phase-space distribution as shown in the
bottom curve of figure \ref{Fig_thermique}. The distribution has a gaussian
shape with a revolution symmetry verified with an agreement better than $1\%$%
. The width of the distribution is
\begin{equation}
\Delta X_{1}=\Delta X_{2}=36.6\;10^{-17}\ m.  \label{Equ_Dxtherm}
\end{equation}

This result can be compared to the theoretical value expected for the
Brownian motion of a harmonic oscillator. The two quadratures $F_{T_{1}}$
and $F_{T_{2}}$ of the Langevin force are uncorrelated and have a flat
spectrum ($S_{T_{1}}=S_{T_{2}}=2S_{T}$). From eqs. (\ref{Equ_X1/F}), (\ref
{Equ_X2/F}) and (\ref{Equ_ST}) one gets the spectrum for the two position
quadratures
\begin{equation}
S_{X_{1}}\left[ \omega \right] =S_{X_{2}}\left[ \omega \right] =\frac{\Gamma
}{M\Omega _{M}^{2}\left( \omega ^{2}+\Gamma ^{2}/4\right) }k_{B}T.
\end{equation}
Both noise spectra are centered around zero frequency with a
Lorentzian shape of width $\Gamma $. Integration over frequency
leads to the variances
\begin{equation}
\Delta X_{1}^{2}=\Delta X_{2}^{2}=\frac{k_{B}T}{M\Omega _{M}^{2}}.
\label{Equ_VarQuad}
\end{equation}
According to the characteristics of the harmonic oscillator (eq.\ \ref
{Equ_Params}) one gets dispersions for the two quadratures of $%
36.3\;10^{-17}\ m$ in excellent agreement with the experimental value (eq.\
\ref{Equ_Dxtherm}).

\begin{figure}[h]
\resizebox{6 cm}{!}{\includegraphics{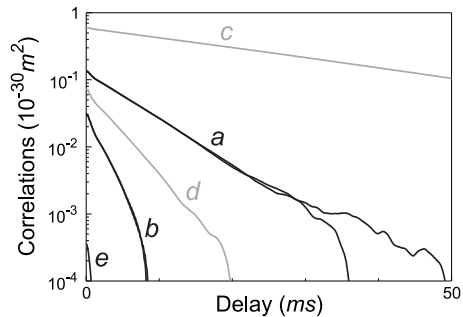}} \caption{Plots in lin-log
scale of the temporal correlation functions $C_{11}\left( \tau \right)$,
$C_{22}\left( \tau \right)$ for both quadratures as a function of the delay
$\tau$. {\it a}: free motion at room temperature. {\it b}: cold damped mirror
(section \ref{Sec_ColdDamping}). {\it c} and {\it d}: parametric cooling
(section \ref{Sec_Param}). {\it e}: phase noise of light.} \label{Fig_Correl}
\end{figure}

Finally we have determined the correlation function of the motion in
phase-space defined for quadratures $X_{i}$, $X_{j}$ ($i,j=1,2$) as
\begin{equation}
C_{ij}\left( \tau \right) =\left\langle X_{i}\left( t\right) X_{j}\left(
t+\tau \right) \right\rangle _{t},  \label{qu_DefC}
\end{equation}
where the brackets $\left\langle ...\right\rangle _{t}$ stand for the
temporal average over the measurement time. Curves {\it a} in figure \ref
{Fig_Correl} show the experimental result obtained for an acquisition time
of $10$ minutes. As expected for a harmonic oscillator in thermal
equilibrium, there is no cross-correlation between the two quadratures ($%
C_{12}=C_{21}=0$) and auto-correlation functions $C_{11}$ and $C_{22}$ are
equal and exhibit an exponential decay from their initial values
corresponding to the variances,
\begin{equation}
C_{ii}\left( \tau \right) =\Delta X_{i}^{2}\exp \left( -\Gamma \tau
/2\right) .  \label{Equ_CThermal}
\end{equation}
The slope of the curve leads to a damping constant $\Gamma /2\pi $ of $48$ $%
Hz$ in good agreement with the expected value ($\Gamma /2\pi \simeq 43$ $Hz$%
).

Curve {\it e} in figure \ref{Fig_Correl} represents the correlation function
for the phase noise of the reflected beam obtained when the cavity is out of
resonance with the incident laser.\ One gets an exponential decay related to
the cut-off frequency of the low-pass filters of the demodulation system.\
This curve clearly shows that the measurement noise is almost negligible as
compared to the thermal noise of the mirror.

\section{Cold-damped regime}

\label{Sec_ColdDamping}

We now study the temporal evolution of the mirror in the
cold-damped regime obtained by freezing the motion with an
additional radiation pressure applied on the mirror. The principle
of the cold damping mechanism \cite{Milatz53} is to monitor the
thermal motion and to use an electronic feedback loop which
corrects the displacement by applying an appropriate damping force
\cite {Cohadon99}.

\begin{figure}[h]
\resizebox{6 cm}{!}{\includegraphics{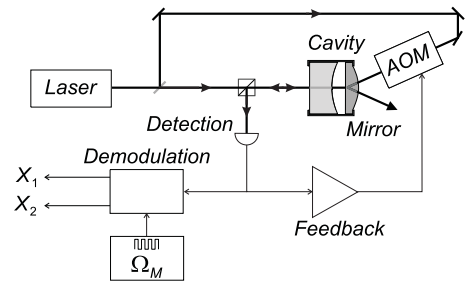}} \caption{Cold-damping
experimental setup. The mirror is submitted to the radiation pressure of an
auxiliary laser beam reflected from the back on the mirror and
intensity-modulated by an acousto-optic modulator (AOM). The feedback loop
allows to apply an additional viscous force.} \label{Fig_Setup_cd}
\end{figure}

As shown in figure \ref{Fig_Setup_cd} an acousto-optic modulator and an
intensity-modulated beam is added to the experimental setup. This allows to
apply a controlled radiation pressure on the mirror. An electronic
differentiator with a variable gain drives the modulator to create a viscous
force, that is a force proportional to the measured velocity of the mirror.
The incident beam on the modulator is intensity-stabilized in order to apply
a well calibrated force.

The effect of this force is to modify the damping of the mirror.
It can be described as a change in the mechanical susceptibility
which now becomes \cite {Cohadon99}
\begin{equation}
\chi _{fb}=\frac{1}{M\left( \Omega _{M}^{2}-\Omega ^{2}+\left( 1+g\right)
\Gamma \Omega \right) },  \label{Equ_chifb}
\end{equation}
where $g$ is a dimensionless parameter characterizing the gain of the
feedback loop and proportional to the electronic gain and to the intensity
of the auxiliary beam. As no additional thermal noise is associated with the
cooling process, the mirror motion is still described by the evolution
equation (\ref{Equ_x}) where the mechanical susceptibility $\chi $ is
replaced by $\chi _{fb}$. The resulting motion is equivalent to a thermal
equilibrium but at a different temperature, depending on the gain $g$ of the
loop.\ The fluctuations-dissipation theorem allows to define an effective
temperature $T_{eff}$ which is smaller than the room temperature $T$ for a
positive gain $g$,
\begin{equation}
T_{eff}=\frac{T}{1+g}.  \label{Equ_Teff}
\end{equation}
The freezing of the mirror of course goes with a reduction of the Brownian
motion which would appear as a shrinking of the distribution in phase-space.
The variances of quadratures $X_{1}$ and $X_{2}$ can be calculated in the
same way as in previous sections and one gets expressions similar to eq. (%
\ref{Equ_VarQuad}) with temperature $T$ replaced by $T_{eff}$,
\begin{equation}
\Delta X_{1}^{2}=\Delta X_{2}^{2}=\frac{k_{B}T}{M\Omega _{M}^{2}\left(
1+g\right) }.  \label{Equ_VarQuadCD}
\end{equation}
The change in the mechanical susceptibility also affects the spectra of $%
X_{1}$ and $X_{2}$. They still have a Lorentzian shape but with a
width equal to the effective damping $\left( 1+g\right) \Gamma $
and a power at resonance reduced by a factor $\left( 1+g\right)
^{2}$.

The cold damping of a mirror has already been demonstrated by
studying the thermal noise spectrum of the mirror
\cite{Cohadon99,Pinard01}.\ Noise reductions larger than $30$ $dB$
at the mechanical resonance frequency and temperature reduction
factors larger than $10$ have been obtained. We
present here the results obtained in phase-space for a moderate gain $%
g\simeq 3$.

\begin{figure}[h]
\resizebox{5 cm}{!}{\includegraphics{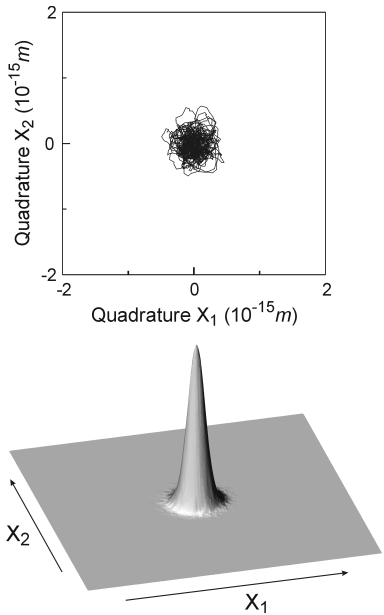}} \caption{Cold-damped
motion with a gain $g=3$, in the $\left(X_1,X_2\right)$ phase-space (same
scale as in figure \ref{Fig_thermique}). The shrinking of the motion is due
to the cooling mechanism. Top: temporal acquisition over $500$ $ms$. Bottom:
corresponding histogram acquired over 10 minutes.} \label{Fig_colddamp}
\end{figure}

The motion of the cold-damped mirror is shown in figure \ref{Fig_colddamp}
(top curve) which also displays the related histogram (bottom curve). As
expected this distribution has a gaussian shape with cylindrical symmetry,
but with a reduced width as compared to the initial thermal distribution of
figure \ref{Fig_thermique}.\ Ratios of the variance of the quadratures
between the two situations are in agreement with eq.\ (\ref{Equ_VarQuadCD})
and with the value of the feedback gain.

Curves {\it b} in figure \ref{Fig_Correl} show the auto-correlation
functions for the two quadratures.\ They correspond to the ones of a
harmonic oscillator in thermal equilibrium at a lower temperature $T_{eff}$%
.\ Compared to the Brownian motion at room temperature, the initial value at
$\tau =0$ is decreased as the variances and the time constant of the
exponential decay is reduced.\ This is associated with the increase of the
effective damping by the cold damping mechanism ($\left( 1+g\right) \Gamma
/2\pi \simeq 170$ $Hz$).

\section{Parametric cooling}

\label{Sec_Param}

Distributions obtained in previous sections are symmetric since there is no
privileged quadrature. We now explore the possibility to obtain asymmetric
distributions, that is to squeeze the thermal noise.\ By analogy with
quantum squeezed states of light, one convenient way to achieve such states
is to use parametric amplification.\ One can take advantage of the presence
of an intensity-controlled auxiliary laser beam to modulate some parameters
of the mechanical oscillator at twice the resonance frequency $\Omega _{M}$.

Parametric amplification of a mechanical harmonic oscillator is
usually done by modulating its spring constant
\cite{Rugar91,Landau60}. An equivalent mechanism is obtained for a
mirror by applying a modulated force proportional to the position
$x$ of the mirror,
\begin{equation}
F\left( t\right) =2gM\Gamma \Omega _{M}\cos \left( 2\Omega _{M}t\right)
x\left( t\right) ,  \label{Equ_FParamX}
\end{equation}
where $g$ is the gain of the parametric amplification. This can easily be
done in our experiment by monitoring the mirror position and accordingly
controlling the force applied by the auxiliary beam.\ Since we use an
electronic control it is even possible to differentiate the signal given by
the homodyne detection as in the case of cold damping, so that we apply a
modulated force proportional to the velocity $\dot{x}$ of the mirror,
\begin{equation}
F\left( t\right) =2gM\Gamma \cos \left( 2\Omega _{M}t\right) \dot{x}\left(
t\right) .  \label{Equ_FParamV}
\end{equation}
This corresponds to the less usual case of a parametric amplification via a
modulation of the relaxation rate of the mechanical oscillator.

For such a viscous force one gets from eqs.\ (\ref{Equ_X1}) and (\ref{Equ_X2}%
) the two quadratures $F_{1}$ and $F_{2}$ of the applied force,
\begin{eqnarray}
F_{1}\left[ \omega \right] &=&gM\Gamma \Omega _{M}X_{2}\left[ \omega \right]
, \\
F_{2}\left[ \omega \right] &=&gM\Gamma \Omega _{M}X_{1}\left[ \omega \right]
.
\end{eqnarray}
According to the evolution equations (\ref{Equ_X1/F}) and (\ref{Equ_X2/F})
one finds that the two quadratures $X_{1}$ and $X_{2}$ of the mirror
position are decoupled and obey the following equations,
\begin{eqnarray}
X_{1}\left[ \omega \right] &=&-\frac{1}{2M\Omega _{M}}\left( \frac{1}{%
-i\omega +\Gamma _{1}/2}\right) F_{T_{2}}\left[ \omega \right] ,
\label{Equ_X1Param} \\
X_{2}\left[ \omega \right] &=&\frac{1}{2M\Omega _{M}}\left( \frac{1}{%
-i\omega +\Gamma _{2}/2}\right) F_{T_{1}}\left[ \omega \right] ,
\label{Equ_X2Param}
\end{eqnarray}
where $F_{T_{1}}$ and $F_{T_{2}}$ are the two quadratures of the Langevin
force $F_{T}$ and where $\Gamma _{1}$, $\Gamma _{2}$ are the
quadrature-dependent effective dampings in presence of the modulated
feedback force,
\begin{eqnarray}
\Gamma _{1} &=&\Gamma \left( 1+g\right) ,  \label{Equ_g1} \\
\Gamma _{2} &=&\Gamma \left( 1-g\right) .  \label{Equ_g2}
\end{eqnarray}

As for the cold damping mechanism, the parametric amplification
changes the damping of the oscillator but the effect now depends
on the quadrature, one damping being increased while the other one
is decreased. As a result, one quadrature of the motion is
amplified whereas the other one is attenuated. From eqs.
(\ref{Equ_X1Param}) and (\ref{Equ_X2Param}) one gets the
variances,
\begin{eqnarray}
\Delta X_{1}^{2} &=&\frac{k_{B}T}{M\Omega _{M}^{2}}\frac{\Gamma }{\Gamma _{1}%
},  \label{Equ_VarQuadParam1} \\
\Delta X_{2}^{2} &=&\frac{k_{B}T}{M\Omega _{M}^{2}}\frac{\Gamma }{\Gamma _{2}%
},  \label{Equ_VarQuadParam2}
\end{eqnarray}
which have to be compared to eq. (\ref{Equ_VarQuad}).

In the case of a modulated restoring force (eq. \ref{Equ_FParamX}) the two
quadratures $F_{1}$ and $F_{2}$ of the force respectively depend on $X_{1}$
and $X_{2}$ so that the evolution equation of $X_{1}$ and $X_{2}$ are no
longer decoupled.\ One has to consider new quadratures obtained by a
rotation of $45%
{{}^\circ}%
$ in phase-space in order to obtain decoupled equations similar to eqs.\ (%
\ref{Equ_X1Param}) and (\ref{Equ_X2Param}), with the same effective dampings
$\Gamma _{1}$ and $\Gamma _{2}$. Both forces thus give the same results
except for a global rotation in phase-space.

\begin{figure}[h]
\resizebox{6 cm}{!}{\includegraphics{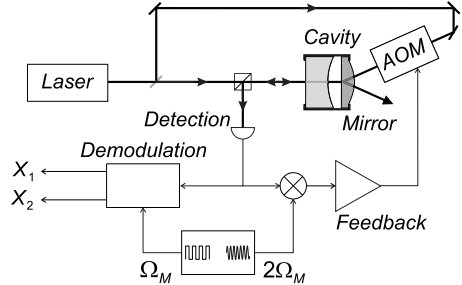}} \caption{Experimental
setup of parametric amplification. An electronic mixer modulates the signal
of the feedback loop at twice the mechanical resonance frequency $\Omega_M$.
The demodulation system is synchronized with the reference signal at
$2\Omega_M$.} \label{Fig_Setup_cp}
\end{figure}

The feedback loop in our experimental setup is modified in order to modulate
the amplitude of the force applied by the auxiliary laser beam at twice the
resonance frequency $\Omega _{M}$ (figure \ref{Fig_Setup_cp}). A reference
signal is used to synchronize this modulation with the demodulation of the
quadratures. We can change the phase between the two signals at $\Omega _{M}$
and $2\Omega _{M}$ in order to select the decoupled quadratures, either $%
X_{1}$ and $X_{2}$ for a modulated viscous force (eq.\ \ref{Equ_FParamV}) or
quadratures rotated by $45%
{{}^\circ}%
$ for a modulated restoring force (eq.\ \ref{Equ_FParamX}).

\begin{figure}[h]
\resizebox{5 cm}{!}{\includegraphics{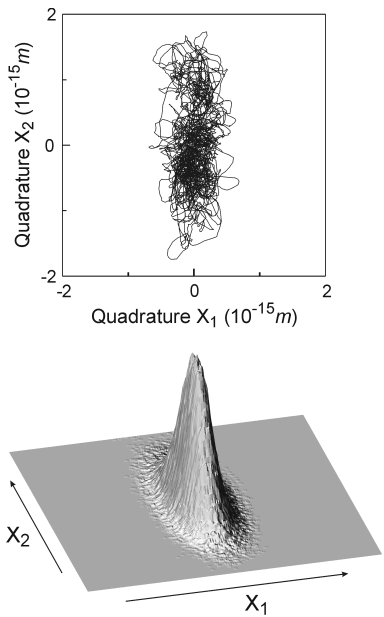}} \caption{Parametric
amplification of the Brownian motion for a gain $g=0.8$, in the
$\left(X_1,X_2\right)$ phase-space (same scale as in figure
\ref{Fig_thermique}). The squeezing of the thermal noise is due to the
quadrature-dependent cooling. Top: temporal acquisition over $500$ $ms$.
Bottom: corresponding histogram acquired over 10 minutes.}
\label{Fig_CoolParam}
\end{figure}

The resulting motion in phase space is shown in figure \ref{Fig_CoolParam}
for a parametric gain $g=0.8$. The distribution still has a gaussian shape
but with different widths for the two quadratures. The distribution is no
longer symmetric and dispersions of both quadratures are found equal to
\begin{equation}
\Delta X_{1}=27\;10^{-17}\ m,\;\Delta X_{2}=78\;10^{-17}\ m.
\label{Equ_DxParamCool}
\end{equation}
Compared to the dispersion obtained at room temperature (eq.\ \ref
{Equ_Dxtherm}) the variance of quadrature $X_{1}$ is reduced by a factor $%
1/\left( 1+g\right) $ whereas the one of quadrature $X_{2}$ is increased by
a factor $1/\left( 1-g\right) $. One quadrature is thus cooled while the
other one is heated and the state corresponds to a thermal squeezed state.

Curves {\it c} and {\it d} of figure \ref{Fig_Correl} show the
auto-correlation functions for the two quadratures $X_{2}$ and $X_{1}$,
respectively. Initial values ($\tau =0$) correspond to the variances deduced
from eq. (\ref{Equ_DxParamCool}) and the time constants of the exponential
decays are related to the effective dampings $\Gamma _{2}$ for curve {\it c}
and $\Gamma _{1}$ for curve {\it d}, in agreement with eqs. (\ref{Equ_g1})
and (\ref{Equ_g2}). Note that we have aligned the two quadratures $X_{1}$
and $X_{2}$ with the axes of the squeezed thermal state by a proper choice
of the phase between the two reference signals at $\Omega _{M}$ and $2\Omega
_{M}$. We have then checked that there is no cross-correlation between the
two quadratures ($C_{12}=C_{21}=0$).

The treatment presented here is based on linear response theory and is only
valid for a gain $g$ smaller than $1$.\ For a gain equal to $1$, the
effective damping $\Gamma _{2}$ vanishes (eq.\ \ref{Equ_g2}) and the
variance of quadrature $X_{2}$ becomes infinite (eq.\ \ref{Equ_VarQuadParam2}%
). This actually corresponds to the oscillation threshold of the parametric
amplification and will be studied in next section.

\begin{figure}[h]
\resizebox{5.8 cm}{!}{\includegraphics{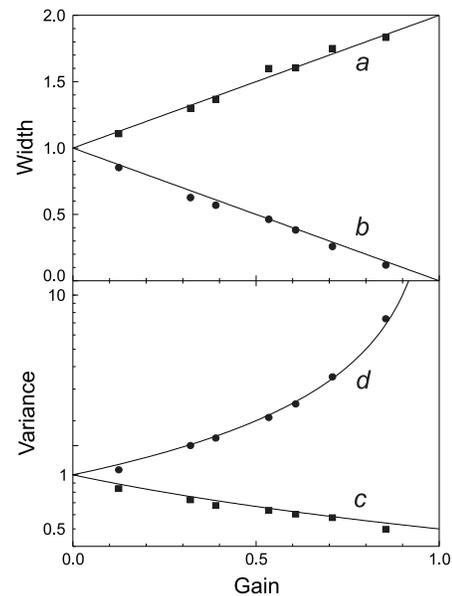}} \caption{Relative values of
effective dampings $\Gamma_1$ ({\it a}), $\Gamma_2$ ({\it b}), and variances
$\Delta X_1^2$ ({\it c}), $\Delta X_2^2$ ({\it d}) normalized to their values
in thermal equilibrium at room temperature, for different gains of
amplification. Solid lines are theoretical predictions.} \label{Fig_Gain}
\end{figure}

We have checked the theoretical behaviour of the parametric amplification as
a function of the gain $g$ in the linear regime ($g<1$). We have repeated
the experiment for different values of the gain and reported the effective
dampings $\Gamma _{i}$ and the variances $\Delta X_{i}^{2}$ for the two
quadratures ($i=1,2$).\ Points in figure \ref{Fig_Gain} show the
experimental results for these parameters, normalized to their respective
values in thermal equilibrium at room temperature. The gain $g$ is estimated
by an average over the independent values of the effective dampings and the
widths of the distributions in phase-space.

The measurements are in very good agreement with the theoretical expressions
given by eqs. (\ref{Equ_g1}) to (\ref{Equ_VarQuadParam2}) and shown as solid
lines in figure \ref{Fig_Gain} without any adjustable parameter. Both
dampings have linear and opposite dependence with the gain. The variance of
quadrature $X_{2}$ increases and diverges at the oscillation threshold $g=1$%
, whereas the one of quadrature $X_{1}$ goes down to $0.5$. The efficiency
of parametric amplification which is mainly related to the power of the
auxiliary laser beam is large enough to obtain a gain $g$ close to unity.\
One then reaches the $50\%$ theoretical limit of squeezing for the variance
of quadrature $X_{1}$.

\section{Parametric oscillation}

\label{Sec_ParamOsc}

As in usual parametric amplification the system oscillates above threshold,
that is for a gain $g$ larger than $1$. This threshold is actually a
consequence of the fact that the gain $g\Gamma $ of parametric amplification
becomes equal to the mechanical losses which are related to the damping $%
\Gamma $. The damping $\Gamma _{2}$ then vanishes and the noise of
quadrature $X_{2}$ becomes infinite. Above threshold the mirror oscillates
at frequency $\Omega _{M}$ in phase with quadrature $X_{2}$. The amplitude
of oscillation depends both on the gain and on saturation mechanisms which
take place in the parametric oscillation regime.\ In our case they are
mainly related to the saturation of the intensity modulation of the
auxiliary laser beam.

The distribution of the mirror motion no longer appears as a peak
located at the center of phase-space ($X_{1}=0$, $X_{2}=0$).\ It
rather looks like a squeezed distribution centered at a non-zero
position along the $X_{2} $ axis, the amount of squeezing
depending on the distance to threshold. As the modulation of the
feedback loop is at frequency $2\Omega _{M}$, the system is
invariant by a $\pi $ phase-shift and there are two possible
positions for the center of the distribution, one for a positive value of $%
\left\langle X_{2}\right\rangle $, the other one for a negative and opposite
value $-\left\langle X_{2}\right\rangle $.

\begin{figure}[h]
\resizebox{5 cm}{!}{\includegraphics{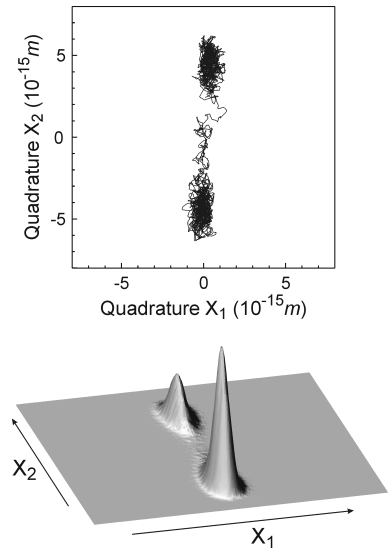}} \caption{Parametric
oscillation of the mirror for a gain $g>1$, in the $\left(X_1,X_2\right)$
phase-space. Scale is $4$ times larger than in figure \ref{Fig_thermique}.
Top: temporal acquisition over $500$ $ms$. Bottom: corresponding histogram
acquired over 10 minutes. The distribution appears as two opposite squeezed
peaks with non-zero equal mean amplitudes.} \label{Fig_CoolAbov}
\end{figure}

\begin{figure}[h]
\resizebox{5 cm}{!}{\includegraphics{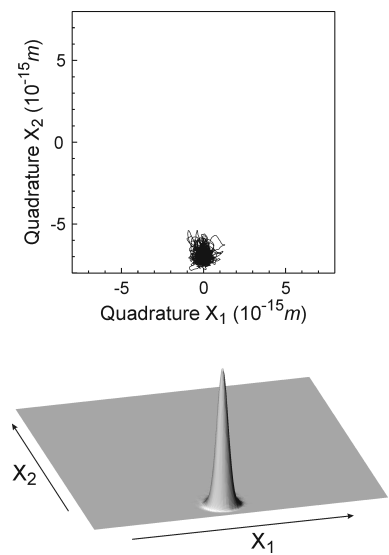}} \caption{Parametric
oscillation of the mirror for a gain $g$ much larger than $1$, in the
$\left(X_1,X_2\right)$ phase-space (same scale as in figure
\ref{Fig_CoolAbov}). Top: temporal acquisition over $500$ $ms$. Bottom:
corresponding histogram acquired over 10 minutes.} \label{Fig_CoolAbo2}
\end{figure}

Figures \ref{Fig_CoolAbov} and \ref{Fig_CoolAbo2} show the
experimental temporal evolutions and distributions obtained above
threshold, respectively for a gain close to unity and for a larger
gain. As expected the distribution in figure \ref{Fig_CoolAbov}
exhibits two peaks along the $X_{2} $ axis, symmetrically located
with respect to the center. The temporal evolution (top curve in
figure \ref{Fig_CoolAbov}) also shows a concentration of the
mirror motion around the two opposite positions.\ One can however
note that the motion sometimes explores the neighbourhood of the
center of phase-space and can jump from one position to the other
one.\ Delay between two jumps can be as small as a few seconds for
a gain very close to unity and corresponds to a few minutes in the
case of figure \ref {Fig_CoolAbov}.\ Over the $10$-minutes
acquisition time of the distribution the system has actually made
only one jump and it has stayed a longer time near the negative
position than near the positive one.\ This explains why the
negative peak is higher than the positive one.\ For a longer
acquisition time one would obtain symmetric peaks.

For a larger gain the system can stay a few hours without any
jump.\ During the acquisition time the mirror then stays near one
position and the distribution exhibits only one peak as shown in
figure \ref{Fig_CoolAbo2}.

Parametric oscillation also squeezes the mirror motion along the
$X_{1}$ quadrature.\ As usual the squeezing effect is larger close
to threshold and the distribution goes back to a disk for large
gain (figure \ref {Fig_CoolAbo2}).\ In contrast with parametric
oscillation in optics \cite {Collett84,Gardiner84}, the
fluctuations are of the same order as the mean value of the
oscillation, that is the distance between the two peaks in figure
\ref{Fig_CoolAbov} is of the same order as their widths. As a
consequence one cannot use a linear approach to describe the
evolution of fluctuations and the distribution does not appear as
an ellipse but as a more complex and distorted shape
\cite{Reynaud92}. It is however possible to evaluate the variance
$\Delta X_{1}^{2}$ and one obtains in the case of figure
\ref{Fig_CoolAbov} a variance equal to $0.55$ times the variance
of the thermal noise at room temperature.

Let us finally note that the amplitude of the peaks (distance to the center
of phase-space) depends on the gain.\ One gets a larger amplitude for a
higher gain.\ There is however a limit related to the saturation of the
feedback loop. For a very large gain the radiation pressure applied on the
mirror is limited by the modulation capabilities of the laser beam
intensity.\ One can easily estimate this limit.\ When the feedback loop
saturates, the intensity modulation tends towards a square signal with a $%
100\%$ modulation depth.\ The component of the radiation pressure force at
frequency $\Omega _{M}$ is thus on the order of
\begin{equation}
F_{rad}\left( t\right) \simeq 2\frac{P}{c}\frac{4}{\pi }\cos \left( \Omega
_{M}t\right) ,
\end{equation}
where $P$ is the light power.\ In this expression $2P/c$ represents the
constant force exerted by light without modulation and $4/\pi $ is a
correction factor due to the fact that the modulation corresponds to a
square signal rather than to a sine one. The response of the mirror is a
forced oscillation with a non-zero mean amplitude $\left\langle
X_{2}\right\rangle $ obtained by identifying the intrinsic damping force to
the radiation pressure,

\begin{equation}
M\Gamma \Omega _{M}\left\langle X_{2}\right\rangle \simeq 2\frac{P}{c}\frac{4%
}{\pi }.
\end{equation}
For a $500\ mW$ beam, this corresponds to a mean displacement of $6$ $%
10^{-15}\ m$, in very good agreement with the one observed in figure \ref
{Fig_CoolAbo2}.

\section{Conclusion}

We have presented an experiment of high-sensitivity displacement
measurement, reconstructing the phase-space distribution of a mechanical
oscillator with a thousand-fold increase in sensitivity upon what had
previously been reported.\ The sensitivity of our experimental setup is
currently limited to $1.7\;10^{-17}\ m$ by the damping timescale of the
oscillator, but it could easily be extended to the attometer level by
reducing the analysis bandwidth.

Both feedback schemes studied in this paper are able to reduce the
thermal noise of the mirror.\ Their effects however are
different.\ In the case of cold damping the resulting state is
still a thermal equilibrium but at a lower effective temperature,
in principle down to a zero temperature for a very large gain
\cite{Courty01}. One then gets a symmetric gaussian distribution
in phase-space.\ In contrast, parametric amplification only cools
one quadrature of the mirror motion, the other one being heated.\
One then gets squeezed thermal states with an asymmetric gaussian
distribution in phase-space.\ This situation is somewhat
equivalent to the one which would be obtained with
quadrature-dependent feedback loops where the zero-temperature
quantum state of the mirror should be reached via squeezed thermal
states \cite{Vitali02}.

These results show that the observation of motion in phase-space gives a
better understanding and control of optomechanical coupling, on the way to
experimental demonstration of related quantum effects.


\begin{thebibliography}{}
\bibitem{CommentDamping}  Damping for internal acoustic modes may not be
viscous, but as we only consider the motion at frequencies close to the
mechanical resonance, there is no observable difference. See P.R. Saulson,
Phys. Rev. D {\bf 42}, 2437 (1990).
\end{thebibliography}
\end{document}